# Eagle Pass, TX: The First American City on the Path of Totality: Organizing Eclipse Party on the Stadium

**Maria D. Kazachenko[1,2] Jorge Perez-Gallego[3] Jennifer Miller[4] Francisco Vielma[5] Mitzi Adams[6] Tishanna Ben[3] Marcel F. Corchado-Albelo[3,7,2] Ryan French[3] Olivia Guerrero-Rish[8] Catarino Morales III[8] Leon Ofman[9,10,11] Evan Pascual[3] Claire L. Raftery[3,12] Jonathan Schiller[2] Dennis Tilipman[3,7,2] John Williams[3]**

[1]U.S. National Science Foundation National Solar Observatory, 3665 Discovery Drive, Boulder, CO 80303, USA, Laboratory for Atmospheric and Space Physics, Boulder, CO 80303, USA,

[2]Department of Astrophysical and Planetary Sciences, University of Colorado Boulder, 2000 Colorado Avenue, Boulder, CO 80305, USA,

[3]U.S. National Science Foundation National Solar Observatory, 3665 Discovery Drive, Boulder, CO 80303, USA,

[4]Sul Ross State University, Alpine, TX, 79832, USA,

[5]Eagle Pass Independent School District, Eagle Pass, TX, 80309, USA,

[6]NASA Marshall Space Flight Center, Huntsville, AL 35812, USA,

[7]Laboratory for Atmospheric and Space Physics, Boulder, CO 80303, USA,

**8**Southwest Texas Junior College, Uvalde, TX 78801,

**9**The Catholic University of America, Washington, DC 20064, USA,

**10**NASA Goddard Space Flight Center, Greenbelt, MD, 20771, USA,

**11**Visiting, Tel Aviv University, Tel Aviv, Israel,

**12**University Corporation for Atmospheric Research, Boulder, CO 80303, USA

**ABSTRACT**

In this paper we share the experience of the U.S. National Science Foundation (NSF) National Solar Observatory (NSO) scientists, educators and public outreach officers organizing an eclipse viewing party at a sports complex stadium on the US/Mexico border in Eagle Pass, TX in collaboration with educators from Eagle Pass and Uvalde areas. We describe reasons we chose Eagle Pass, contacts we established with the local community, preparations for and activities set up during the eclipse viewing party, the eclipse day on April 8 2024 and lessons learned from organizing our event.

# Introduction

*Overview*: A few days prior to the April 8th 2024 eclipse, scientists and education officers from the U.S. National Science Foundation (NSF) National Solar Observatory (NSO; from Boulder, CO and Maui, HI), and NASA's Marshall Space Flight Center (MSFC) traveled to San Antonio, TX. From there we continued to locations close to the Mexican border, to lead outreach activities in K-12 schools in the Del Rio, Eagle Pass, and Uvalde school districts for 2-3 days before the eclipse. On the day of the eclipse, we set up activities at the Eagle Pass Student Activity Center (SAC) football stadium, welcoming students and the public from surrounding areas to experience the eclipse with us. We were supported by up to 33 undergraduate student volunteers from both Sul Ross State University's (SRSU) Robert Noyce En La Frontera Scholarship Program and Southwest Texas Junior College (SWTJC) in Uvalde, building upon relationships established ahead of and during the October 2023 annular eclipse in Uvalde, TX. In addition, NASA/MSFC teams set up on the campus of Southwest Texas Junior College in Uvalde and at Garner State Park close by.  In this paper we share our experience organizing eclipse viewing parties, specifically at the Eagle Pass SAC football stadium: reasons we chose Eagle Pass, partnerships with the local community, and preparations for and the main event on eclipse day.

*Project Inception:* The idea for the NSO/CU Boulder team to go to Texas for totality came up in early 2019. At that time Prof. Maria Kazachenko and then head of NSO's Education, Public Outreach and Communications (EPOC) department, Dr. Claire Raftery decided to team up for the education and outreach component of the NSF CAREER proposal of Prof. Kazachenko. Our initial plan was to perform outreach during three solar eclipses: a total solar eclipse in Chile (2020) and an annular and total solar eclipses in Southwest Texas (2023, 2024).  Specifically, we planned to work with the same group of bilingual Hispanic K-12 students from Southwest Texas to provide public outreach activities in the path of totality during these three eclipses. Due to the pandemic we had to cancel our trip to Chile and expand our outreach in Southwest Texas: instead of focusing just on outreach at schools, we did outreach both at schools and for local communities at local STEM and community conferences and watching parties during both the annular and the total eclipses.

*Our Team:* Our team primarily consisted of three groups led by Prof. Maria Kazachenko at CU Boulder/NSO (see Figure 1): (1) solar scientists and educators from CU Boulder/NSO, (2) the EPOC team at NSO, and (3) educators and 33 volunteers from the Uvalde/Eagle Pass/Del Rio areas. The group of scientists/educators from CU Boulder included Dennis Tilipman, Marcel Corchado-Albelo, Dr. Ryan French, Jonathan Schiller (from Fiske Planetarium) and Prof. Maria D. Kazachenko. The EPOC group included Evan Pascual, John Williams, Tishanna Ben, its current head Dr. Jorge Perez-Gallego, Shari Lifson, and its former head Dr. Claire Raftery. The Texas team included Dr. Jennifer Miller-Ray, Olivia L. Guerrero-Rish, Francisco Vielma, Catarino Morales and over 33 volunteers from SRSU and SWTJC, with additional volunteer scientists joining at the event.

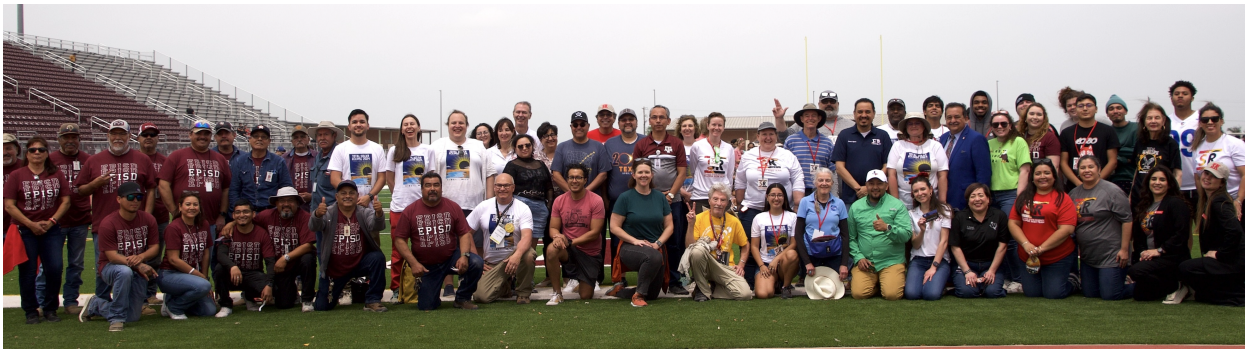

**Figure 1**
*Our eclipse watch party was a team effort between solar scientists from CU Boulder and NSO; NSO's EPOC team in Colorado and Hawaii; NASA's Marshall Space Flight Center; and educators, volunteers and stadium workers from SRSU, SWTJC, Eagle Pass School District, and The City of Eagle Pass, TX. See Introduction for details.*

## Reasons to organize a public event in Eagle Pass, TX

*Initial Plans to go to Uvalde, TX:* As we were planning our outreach activities during the annular and total solar eclipses back in 2021, our initial plan was to do outreach during both events in Uvalde, TX. We chose Uvalde as it had one of the best weather forecasts and lied on the intersection of the paths of the total and annular eclipses. However, as we were preparing for the annular eclipse in Uvalde, and had more interaction with local communities, we were asked to organize our subsequent outreach activities before and during the total eclipse even further away from the big cities, where few educators travel to, in Eagle Pass, TX.

*Why Eagle Pass?* Eagle Pass, TX is a city on the US-Mexico border with a population of around thirty thousand people, which is 95% Hispanic/Latino. We had multiple reasons to choose Eagle Pass. First, Eagle Pass was the first city in the U.S. to experience totality. Second, it had the best historical record for clear skies in April in the U.S. Third, it is remote, with the closest city, San Antonio, located 2.5 hours away by car. To our knowledge at the time, no educators planned to travel to Eagle Pass during the eclipse due to its remoteness, allowing us to maximize the outreach impact of our effort. Fourth, the Kickapoo Traditional Tribe of Texas (KTTT), formerly known as the Texas Band of Traditional Kickapoo, is also located outside of Eagle Pass and is one of three federally recognized Tribes of Kickapoo communities. Across the border, there is an even larger sister Kickapoo community and outreach activities attempting to bridge under-served communities' access to scientists, mentors, and programs to inspire the next generation of STEM leaders. Finally, the Eagle Pass community consists of predominantly underrepresented groups in STEM.

# Preparing for the April 8 2024 Total Solar Eclipse Viewing Party

Here we describe the details of our preparations for the eclipse-related activities in Eagle Pass, TX: establishing new partnerships, developing a timeline, event planning, and funding and implementing eclipse activities.

## Partnerships

*How did we go from zero connections with the local community back in mid 2022 to the eclipse party at a football stadium with the mayor of Eagle Pass and university leaders inaugurating our event?* Back in 2018, as we were writing an NSF CAREER proposal and planning outreach events during the annular and total solar eclipses somewhere in TX, two of us, Prof. Maria Kazachenko and the former head of NSO's EPOC, Dr. Claire Raftery, took a map of TX and created a list of Universities and Community Colleges on the eclipse path. The major institutions that we identified in the area were SWTJC and SRSU. After a few emails, we heard back from Ms. Olivia Guerrero-Rish at SWTJC, who was very enthusiastic about collaborating to organize a series of pre-eclipse outreach events in Uvalde, TX. Ms. Guerrero-Rish later introduced us to Prof. Jennifer Miller-Ray at SRSU. At that time Prof. Miller-Ray was already leading a series of outreach initiatives with support from the National Science Foundation's Noyce en la Frontera program in the area from a mobile STEM Van to Uvalde, Del Rio, and Eagle Pass communities to increase participation in STEM education and to provide mentoring opportunities to SWTJC and SRSU faculty. Programming was based upon work supported in part by the National Science Foundation under Grant 2050173 (Miller, Qvarnstrom, Brown, et. al, 2021). Later, Prof. Jennifer Miller-Ray introduced us to other partners, who made our effort possible: Science & Social Studies Director at Eagle Pass Independent School District Mr. Francisco Vielma who helped us coordinating our educational efforts with the local school district and logistics at the SAC school stadium event; Mr. Ronny Rivera and Ms. Aide Castaño at the City of Eagle Pass who helped to coordinate our event with the city of Eagle Pass and University leadership of the SWTJC and SRSU. All through the planning phase of our outreach activities, we had immense support from the local community.

## Eclipse Planning Timeline

Below we provide a more detailed timeline of our eclipse planning:

- *July 2019:* Prof. Kazachenko wrote an NSF CAREER proposal to organize a series of eclipse-centered outreach events in Chile (2020) and Texas (2023 and 2024).

- *November 2019:* NSF CAREER proposal selected.
- *May 2020:* Initial plans to go to Chile for a total solar eclipse were canceled due to pandemic.
- *September 2022:* Started looking for partners in Texas for the annular and total solar eclipses.
- *October 2022:* Identified the first major partner for the annular solar eclipse, Ms. Olivia Guerrero-Rish.
- *March 2023:* Identified the second major partner for the annular solar eclipse, Dr. Jennifer Miller.
- *March 2023-October 2023:* Planned outreach activities before/during the annular solar eclipse in Uvalde, TX.
- *November 2023-April 2024:* Planned outreach activities before/during the total solar eclipse in Eagle Pass, TX.

## Events We Participated In and Organized to Prepare Our Outreach Activities

In preparation for our eclipse party we participated in outreach conferences and organized teachers training in the area as described below.

- *AAS Solar Eclipse Planning Workshop, October 2022 Rochester, NY:* In October 2022 Dr. Ryan French and Marcel Corchado-Albelo participated in the [AAS Solar Eclipse Planning Workshop](). The meeting introduced Marcel and Ryan to key players in the U.S. eclipse educator community, and provided training on how to communicate the eclipse experience with the public and K-12 students.
- *Science of STEM Literacy Conference, July 2023, Uvalde, TX:* On July 10 2023 several of us, including Prof. Jennifer Miller, Prof. Maria Kazachenko and Ms. Olivia L. Guerrero-Rish, participated in the [Science of STEM Literacy Conference]() at Sul Ross State University in Uvalde, Texas. Prof. Kazachenko gave a talk on "What do we see during solar eclipses? Scientist's perspective" and answered questions about solar science and solar eclipses. 115 educators attended this conference. Dr. Angela Speck, chair of the physics department at University of Texas San Antonio and of the AAS eclipse task force, provided a closing keynote on eclipse science. Participants reported gains in understanding eclipse science and began to engage in participatory science eclipse programs to include NASA Soundscapes, Citizen CATE, and the eclipse ambassador program.
- *Let's Go Solar Middle School STEM camp, July 2023, Uvalde, TX:* Dr. Jennifer Miller and Catarino Morales, assisted by Noyce scholars and SWTJC STEM club students, led a STEM camp for Uvalde CISD middle school students.
- *Teacher Training, September 2023, Online:* On September 22 2023, we organized an [online teacher training]() for K-12 teachers in the Uvalde, Del Rio and Eagle Pass area on the nature phenomena observed during annular and total solar eclipses.
- *Annular Solar Eclipse Outreach Trip, October 12-15, Uvalde, TX:* The major event before our totality trip was the outreach trip we did for the annular eclipse to Uvalde, TX. During this trip we visited two schools in Uvalde, TX and led outreach during the annular solar eclipse on the SRSU stadium.

- *Total Eclipse Volunteers Training, March 2024, Online*: Two weeks before the eclipse we organized a [zoom training session](#) with our TX volunteer-educators to prepare them for the outreach activities during our eclipse party.

## Equipment and Planned Eclipse Activities

Below we describe our preparations designing and applying for funds for equipment and planned eclipse activities: the Solar Tent we built as part of the AAS Jay M. Pasachoff solar eclipse Mini-grant, the SPD/AAS outreach grant on "Many Ways to View the Sun", nine NSO outreach booths and activities, our telescopes, the STEM Van, eclipse T-shirts, presentations for the Jumbotron and the Sun-centered eclipse music playlist that was played on the stadium.

### AAS Jay M. Pasachoff Solar Eclipse Mini-Grant: The Solar Tent

In preparation for the eclipse in 2023 we applied for the grant from the "AAS Call: Engaging the Public with the April 2024 Solar Eclipse" to build the Sun Science Tent with Dr. Ryan French as a PI.

*The Sun Science Tent* provided a unique sensory and educational experience for the public and K-12 students witnessing the solar eclipse, at the National Solar Observatory's eclipse viewing site at the Eagle Pass SAC stadium in Eagle Pass, TX. We utilized experiments at the intersection between eclipse/Sun science and the K-12 school curriculum, including the investigation of spectroscopy and multi-wavelength nature of light with spectral discharge tubes and infrared experiments. Sun Science Tent also attempted to provide (but limited by cloud) a sensory experience for visitors to experience the partial eclipse phases in a unique and exciting way, creating an immersive eclipse viewing environment using reflective disco balls, sun catchers, solar filter windows, and pinhole cameras. The proposal was selected in January 2024.

### SPD/AAS Outreach Grant on Many Ways to View the Sun

Prior to the eclipse we also applied for and received an outreach grant "Eclipses en la Frontera: Many ways of Observing the Eclipse on the US-Mexico border" from the Solar Physics Division of American Astronomical Society Education and Public Outreach Office. This grant has allowed us to purchase the eclipse-viewing equipment to observe the eclipse with higher spatial resolution for a much larger number of people beyond eclipse glasses. Specifically, we purchased a range of instruments from a sunspotter, 8 pairs of binoculars, a welding mask (fun for kids) and a visitor's photo station to take pictures of the eclipse with a phone.

### NSO outreach booths and activities

In addition to the Solar Tent and the eclipse-viewing equipment to observe the eclipse, we planned to set up the following 9 activity tables previously created and widely used by the NSO-s EPOC team: a solar convection demo, polarization of light demo, ultraviolet beads bracelet station, ask-a-solar-scientist booth (led by Dr. Leon Ofman from NASA Goddard Space Flight Center), make-a-pinhole-viewer station, predict-the-solar-corona-shape booth, a station with hands-on experiments with magnets and iron filings to understand the magnetic sun, moon phases hands-on experiment and a Sun photo booth.

## Telescopes

We had a range of solar telescopes and sunspotters to observe the Sun during partial phases. We had a Coronado SolarMax II 90 Double Stack Solar Telescope w/BF15, two sunspotters, two Coronado $H_{alpha}$ Personal Solar Telescopes, a Celestron Nexstar 8SE and a Meade SCT LX10 EMC. the Sunspotters that we had were folded-path Keplerian Solar Telescopes used to observe sunspots and partial eclipse phases of the eclipse. These allowed larger groups to appreciate the Sunspots and provide younger children the opportunity to draw the partial eclipse. One sunspotter was planned to be used outside the Sun Science Tent  while the second sunspotter was located next to all the telescopes.

## AAS Jay M. Pasachoff Solar Eclipse Mini-Grant: The Mobile STEM Van

Following the October eclipse, Dr. Jennifer Miller received additional support from AAS's Eclipse en la Frontera mobile program to further support mobile outreach with teachers and graduate students to include Noyce scholars and mentors participating in community eclipse science programming. This afforded SRSU the opportunity to further provide outreach to rural communities along the path of totality in Feb-April leading up to the event. Efforts to integrate training materials in university coursework included training around 60 educators and graduate students about eclipse science prior to visiting schools and public libraries. The following mobile pop-up activities were customized utilizing Miller, Thomas, & Maryboy's (2018) eclipse makerspace activities incorporated with Little Singer Navajo School near Winslow, Arizona prior to the 2017 eclipse event in which pop-up activities encourage participatory science serving 1 of 4 career roles:  engineer, natural or social scientist, artist, and space weather journalist.

- Space Weather Journalist:   Learning about space weather and its impact on earth's climate, capturing real data and reporting findings.
-  Solar Scientist:  Students learn about Citizen CATE from SRSU's Noyce scholar and Citizen CATE Southwest Texas lead trainer, Andrea Rivera and learn how to track the Sun and utilize a telescope and a sunspotter.
- Sun Stories:  Making artistic sun stories, songs, coloring books, and creative products to inform the public about layers of the Sun, CME's, solar storms, and culture to celebrate the Sun.
- Engineer:  Making solar glasses and learning about eclipse viewing safety.

Other activities included a read aloud to primary students, NSO's yardstick activity to make a replica of an eclipse, NASA's soundscapes, and modeling using clay and crafts to sing stories about our sun.  The mobile STEM lab shown in Figure 2 served over 7,136  participants leading up to eclipse day and visited schools and public libraries in Hondo, San Antonio, Knippa, Crystal City, Carrizo Springs, Alpine, Presidio, El Dorado,

Kerrville, Johnson City, Uvalde, Del Rio, and Eagle Pass communities providing eclipse glasses and coloring books to all public schools in communities mentioned above.

SRSU education students read to elementary students from mobile lab March 2024.

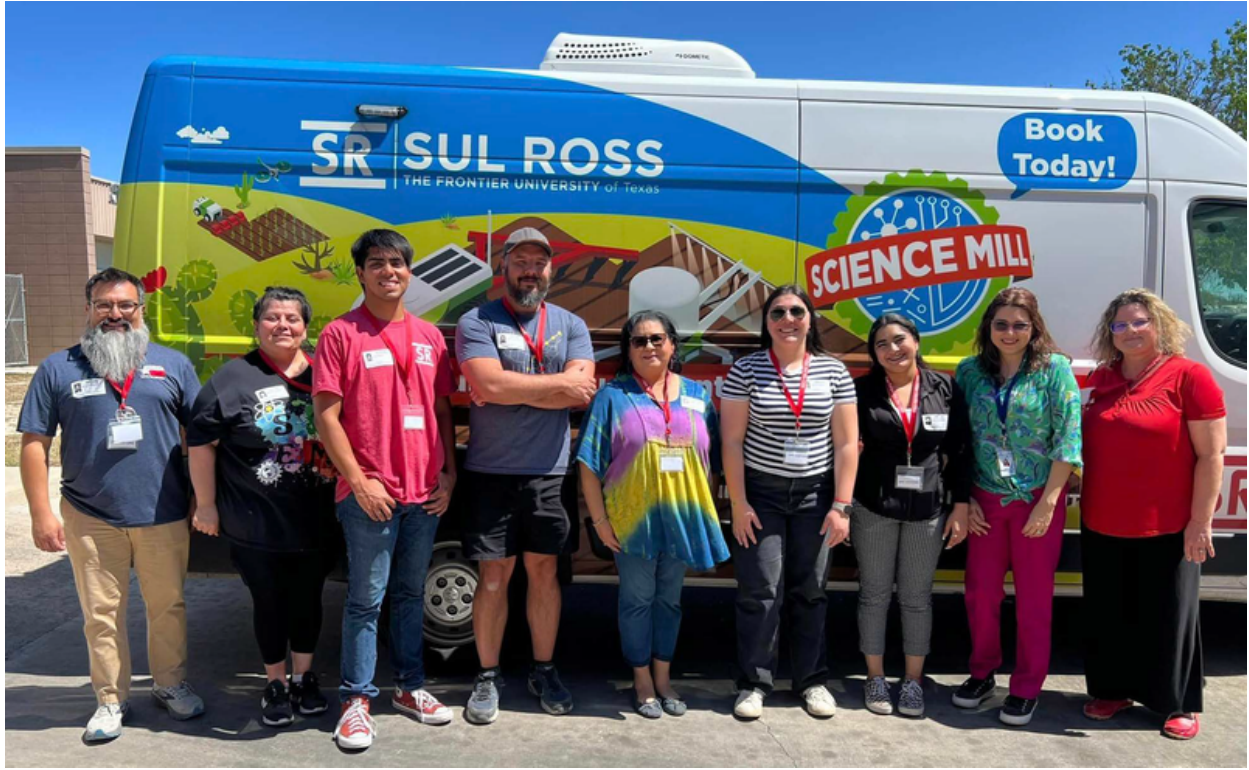

**Figure 2**
Mobile Stem Van: SRSU students, SWTJC Professor Catarino Morales, and SRSU's Dr. Jennifer Miller-Ray with a mobile STEM lab visiting Del Rio CISD elementary students, March 2024.

## Eclipse T-shirts for the Team: Designing and Printing

In preparation for the eclipse event at the stadium we asked the local community at SRSU to design a T-shirt that we would then print on the front of our "Eclipse Ambassador" T-shirt.

Robert Greeson, a Media Specialist at Sul Ross State University and Citizen CATE Eagle Pass team member created a wonderful design shown in Figure 3. We then contacted Apollo Ink Screen Printing in Boulder, CO to print 38 T-shirts for each member of our team. We then wore these T-shirts during our eclipse party. During the event we received lots of compliments and requests on "where-to-buy" our T-shirts. For the future eclipse parties, we definitely recommend printing and selling eclipse T-shirts with custom-made design.

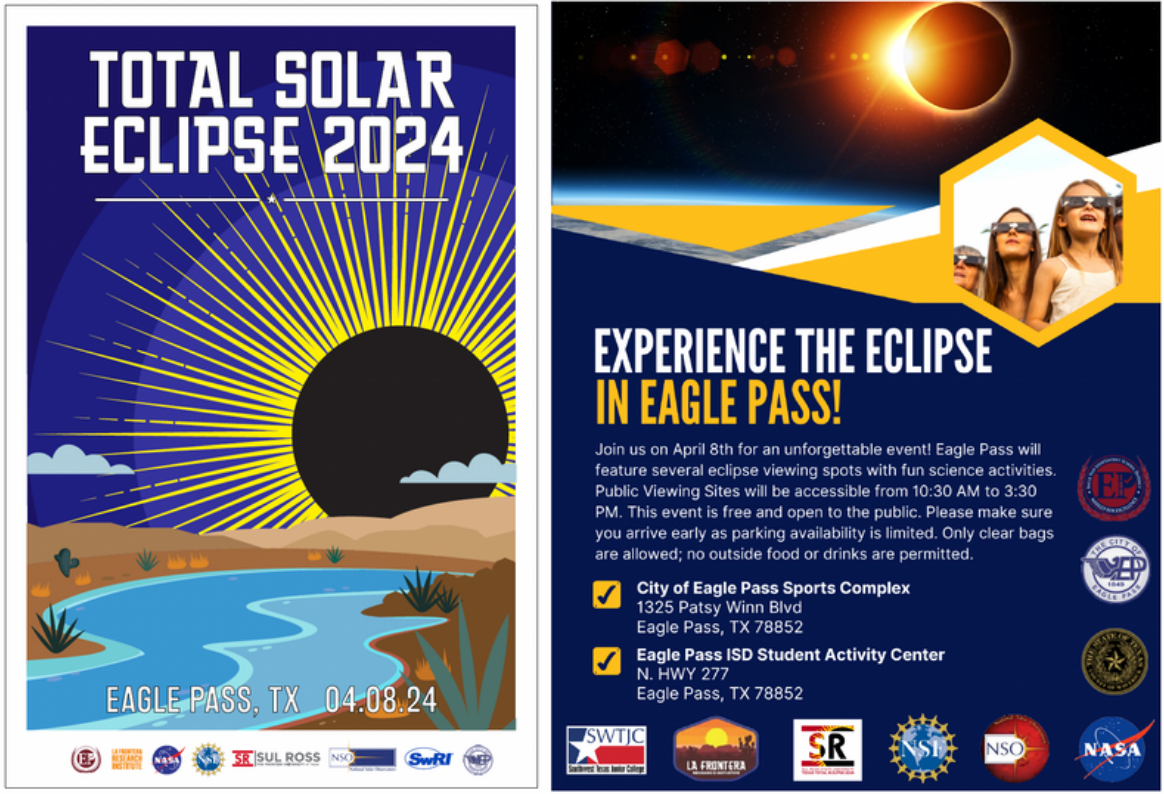

**Figure 3**

Preparations:  Left: A design of the solar eclipse poster and T-shirt made by Robert Greeson at SRSU was used for team members' T-shirts. Right: A flier advertising our eclipse party.

## Solar Fun Facts Slideshow on the Jumbotron

At the SAC stadium we had a jumbotron (47'6.4" high x 56'8.8" wide) where we planned to share fun facts about the Sun before and after the totality (see Figure 5). So we created a presentation "Eclipses en la Frontera: Welcome to the First Eagle Pass Eclipse Viewing Party! The Sun. The Moon. You!" with around 60 slides which we played throughout the event. This presentation included curious facts about the Sun, eclipses, missions studying the Sun, recent discoveries as well as trivia questions to keep the audience engaged.

## The Total Solar Eclipse Music Playlist

The two keys for a successful eclipse party are clear skies and good music. In preparation for the eclipse Dr. Ryan French assembled a crowd-sourced (via X) list of 123 songs on Sun-related subjects in English and Spanish. The playlist is available [on Spotify](on Spotify) and includes a good variety of Sun-centered songs in English and Spanish which we encourage to use during upcoming eclipse parties.

## The Eclipse Day

Below we describe the details of our eclipse day: a detailed schedule of planned events, the weather overview including pre-eclipse clouds and clear skies during totality, performed eclipse activities, attendance and press coverage.

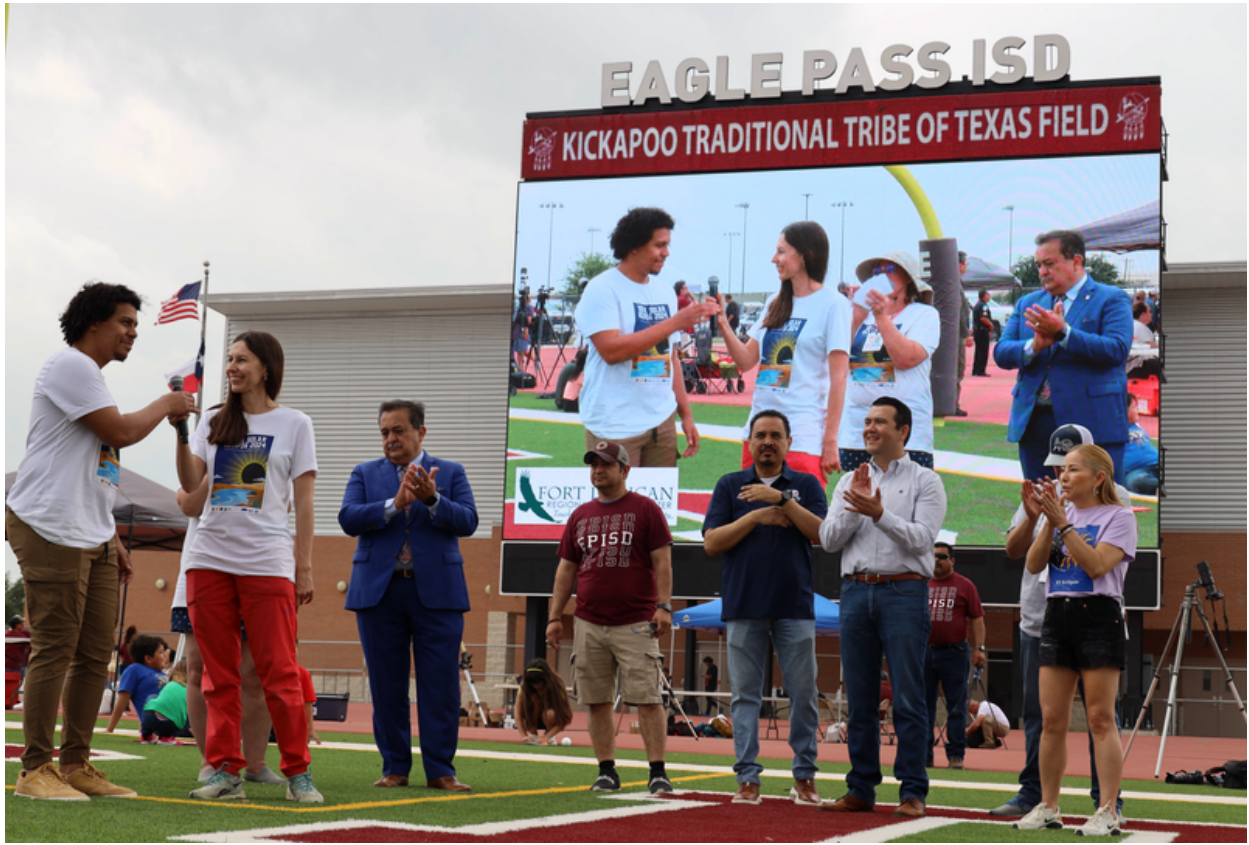

**Figure 4**
Inauguration of our Eclipse Party: from left to right, CU Boulder/NSO scientists Marcel Corchado-Albelo and Prof. Maria Kazachenko, SRSU Prof. Jennifer Miller (behind Maria), SRSU university president, Carlos Hernandez, Eagle Pass ISD School Board President Jorge Barrera, Eagle PASS ISD Superintendent Samuel Mijares, city of Eagle Pass mayor Rolando Salinas and city of Eagle Pass marketing director Aide Castaño.

### Schedule

At 6:30 am our team and the SAC stadium team started setting up the tables. At 10 am the doors opened and we started playing the Sun-inspired music and the slideshow on the jumbotron.  At 12:10 pm the partial phase of the eclipse started and we did a quick welcome of the city to all the attendants (see Figure 4). The welcome included greetings from the Mayor of Eagle Pass, the president of the Sul Ross University,  and description of the events from the NSO team by Prof. Maria D. Kazachenko and Marcel Corchado-Albelo. Three minutes before the totality we stopped the music and the slides to prepare to experience the magic of the totality. We then resumed the music and the jumbotron till the end of the partial phase at 2:51 pm.

## Pre-eclipse clouds

No good story lacks drama, and indeed, the days and minutes leading to the eclipse were filled with suspense. A blanket of clouds swept east across Mexico and Texas, threatening to block the eclipse and disrupt over a year's worth of planning. An entire football stadium slated to be a community watch party filled with telescopes and activities could now face a low turnout instead of the hundreds of attendees expected to arrive on Eclipse Day. The morning sky was soaked in with little promise of seeing the eclipse. But a trickle of attendees soon turned into large crowds, pockets of blue sky became visible, and a glimmer of hope flickered as first contact began.

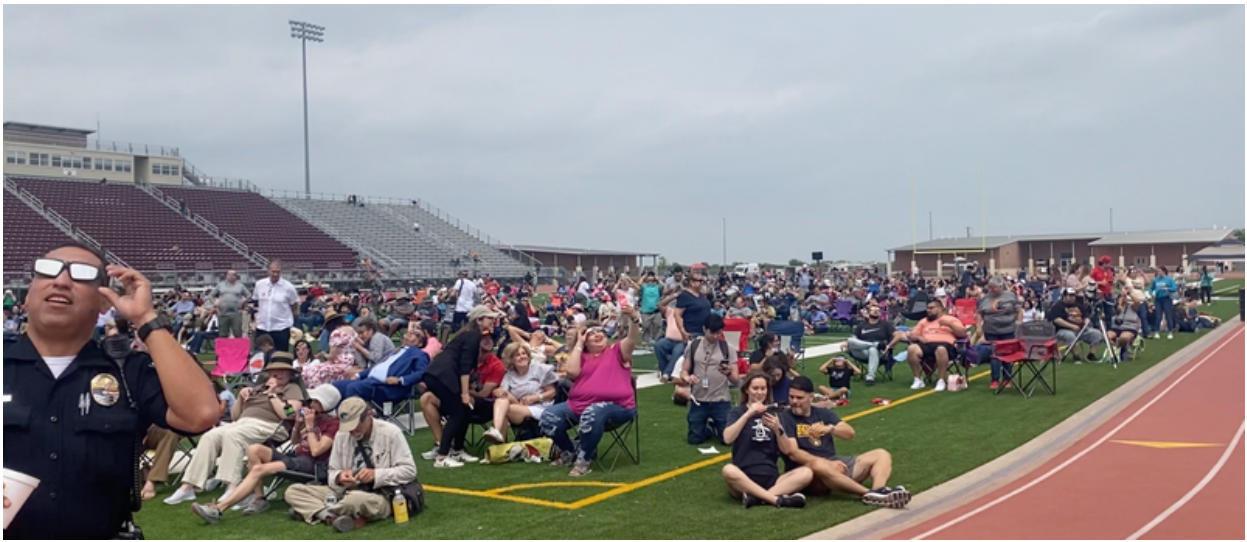

**Figure 6**
Eclipse day: Audience observing the Sun during rare patches of clear sky during the partial phase and learning about the Solar Fun Facts on the stadium jumbotron.

## *Clear skies during totality*

Between the veil of clouds, the partial eclipse pierced through, re-igniting hope for totality. And as luck would have it, for 4 minutes and 24 seconds, the clouds would part just enough to deliver the moment they all been waiting for: a spectacular, awe-inspiring moment in time under the shadow of the moon and Sun. People cried. Some were speechless while others could hardly contain themselves. Children screamed and hollered. Loved ones embraced. Scientists, volunteers, and event staff paused from their duties to lay in the grass and enjoy "the moment".

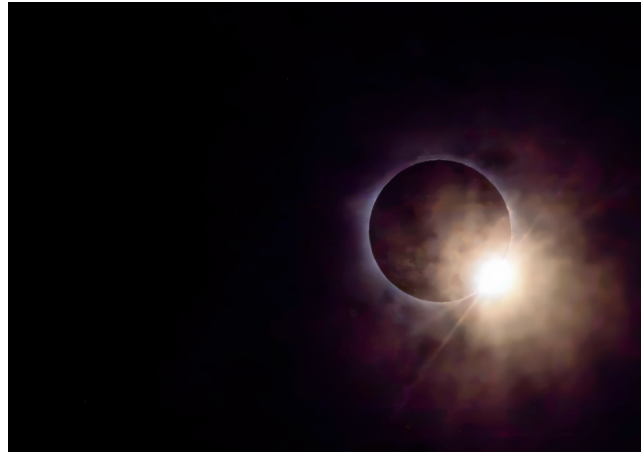

**Figure 7**
In spite of the clouds we were lucky to observe totality with some haze. Credit: John Williams.

## Eclipse Activities

Due to clouds and high winds we had to make on-the-fly changes to our activities. The side walls of the Sun Tent had to be taken off to avoid the tent flying away. All the experiments relying on clear skies and partial eclipse phases, such as pinhole camera, sunspotter, disco ball etc., could not be used, due to very short patches of clear skies. Nevertheless all our demos that did not rely on the clear skies worked well.

## Attendance

We estimate that about 1000 people attended our eclipse party with 119 people answering our questionnaire about their eclipse experience (see Figure 6).  Out of questionnaire respondents 71% were women, 24% were men and 5% did not answer; in terms of ethnic affiliation, 54% were of Hispanic, 40% were White and 6% of other or non-indicate origin. We asked the attendees to indicate the most interesting things they learnt during the event. Here are some representative answers: "how the electrons move around the eclipse at its totality", "learning about the different bubbling cell patterns on the Sun", "the happy environment", "that each cell on the Sun is the size of Texas", "it gets super windy before the eclipse", "the shininess of the solar crown", "the whole experience was just amazing",  "how the eclipse changed the environment. The shadows changed, we could see directly at the Sun... it all felt spiritual!", "how everything went dark and literally everything froze ", "science brings people together", "I was reminded how tiny we are in this vast Universe".

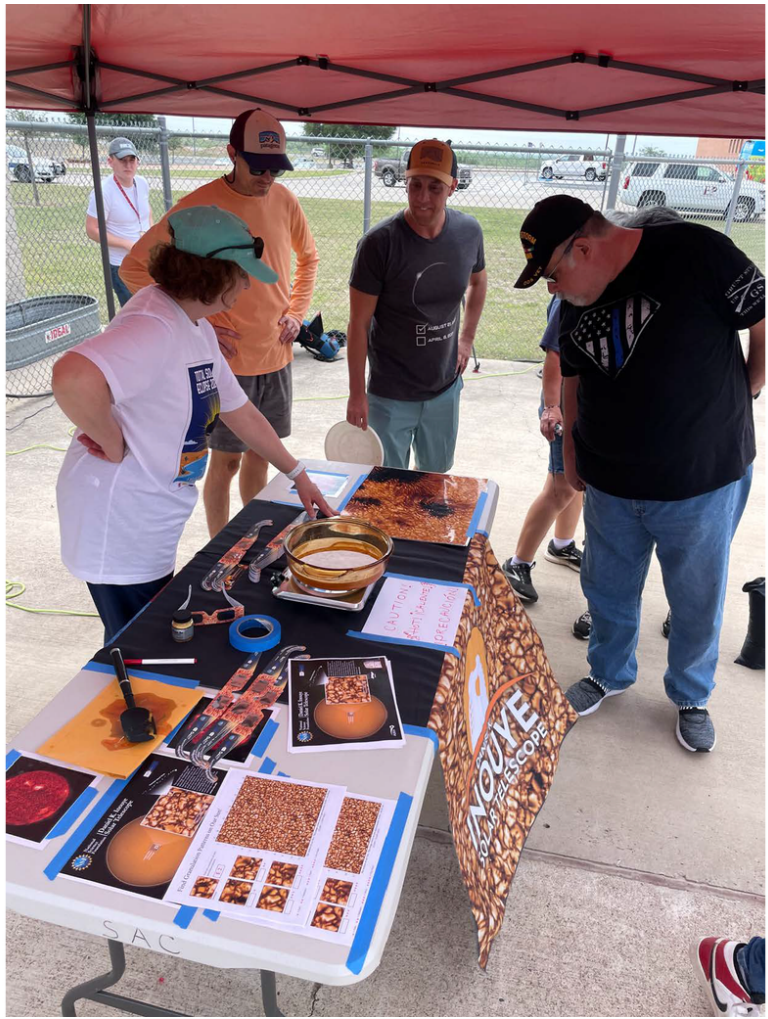

**Figure 8**
One of the NSO EPO demos with hot oil convection cells led by Shari Lifson. On the Sun each cell like that has roughly the size of Texas.

## *Press Coverage*

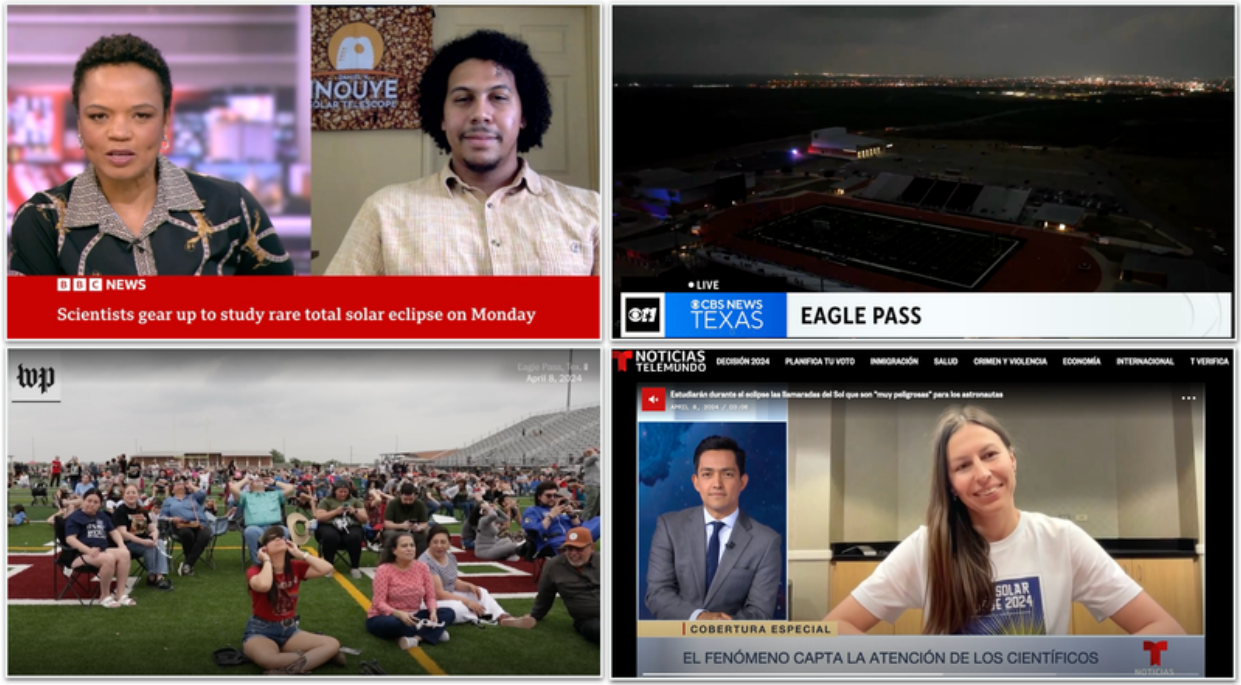

**Figure 9**

Press: Our event and planning activities received good press coverage ranging from BBC News (featuring Marcel Corchado-Albelo), CBS News and Washington Post in English to Telemundo (featuring Maria Kazachenko) in Spanish.

Our eclipse party and the events leading to it received a wide press coverage in over 30 live news reports, blog posts and journal articles. In Figure 9 we show some highlights, which include Washington Post, BBC, Forbes, USA Today, Border Report, Telemundo, Nature etc. In addition, two members of our team, Dr. Jorge Perez-Gallego and Dr. Ryan French, participated in all planning of our eclipse activities, but had to leave the stadium on the day of the eclipse to participate on an [NSF eclipse live stream](#) led by Dr. Perez-Gallego, and co-hosted by Dr. French.

# Conclusions and Lessons Learned

Here we summarize our experience organizing an eclipse viewing party at the stadium.

From finding partners to purchasing new equipment and training volunteers  it took us over a year to plan our eclipse party at the football stadium in Eagle Pass. In preparation for the event we have assembled a team of 33 educators/volunteers, participated in and organized several outreach events and teacher trainings, received several grants from SPD and AAS to purchase equipment, set up over 10 eclipse-event outreach booths and created eclipse-focused science slideshow and music playlist. On eclipse day the weather was far from ideal: we had high winds and thin clouds during partial phases and clear skies during totality. Because of the weather, we could not set up most of our outreach activities that relied on the clear skies, but could still lead many other activities, such as the Solar Fun Facts slideshow on the Jumbotron, solar convection demo or as Ask-a-solar-scientist table. During totality we were lucky to see the corona with light haze from the clouds. Our event received wide press coverage in over 30 news reports. We estimate that about 1000 people attended our event (see Figure 6).

## Lessons learnt

Below we add some takeaways from our experience organizing the event, which we hope might be useful for future planning of eclipse parties.

- Start looking for partnerships and planning early, at least a year before the event.
- Advertise your event to maximize the outreach impact.
- Predicting the number of attendants is a hard task, so prepare to be flexible.
- Plan how to interact with the press on the stadium without sound interference from the stadium sound system.
- Plan activities for cloudy/windy weather.
- Plan to have activities *before* totality rather than after it, as many people tend to leave after totality.
- Most people's attention would be on the Sun. While educational activities are important you could as well keep it simple, having only music and an interesting slideshow with detailed description on what to expect during the eclipse.
- Prepare eclipse-related merchandise to share with the public: eclipse glasses and eclipse T-shirts with custom event design.
- Practice! Practice setting up the scopes, tracking the Sun, preparing the best view of the Sun for visitors and, most importantly, automating as much as possible so that "the moment" can be enjoyed.
- Delegate planned activities to as many people as possible and as early as possible.

## Acknowledgments

M.D.K. acknowledges support from NSF CAREER SPVKK1RC2MZ3 award, Jay Pasachoff Mini-Grant, SPD/AAS Education outreach grant. RF thanks the Brinson Prize Fellowship for supporting scientific research and outreach activities for the 2023 and 2024 eclipses. NSO acknowledges invaluable support from NSF. L.O. acknowledges support from NSF grant AGS-2300961 and NASA GSFC through Cooperative Agreement 80NSSC21M0180 to Catholic University of America, Partnership for Heliophysics and Space Environment Research (PHaSER). SRSU alcknowledges support from NSF's Noyce en la Frontera 2050173 and Jay Paschoff Eclipses en la Frontera Mini-Grant program. M.A. acknowledges assistance from Dr. Adam Kobelski (NASA/MSFC), Dr. Pete Robertson (NASA/MSFC emeritus), Dr. Paul Bremner (NASA/MSFC), Amanda Adams (NASA/MSFC Transform to Open Science (TOPS)), Adam Farragut (NASA/MSFC/TOPS), Derek Wallentinsen (Solar System Ambassadors). We also thank Sul Ross State University and Southwest Texas

Junior College for their support during the eclipse, and the city of Eagle Pass, TX (Rolando Salinas Jr., Mayor) for providing logistic support and security for the event.